\documentclass[aps,prd,preprint,nofootinbib,11pt]{revtex4}
\usepackage{amsfonts}
\usepackage{mathrsfs}
\usepackage{graphicx}
\usepackage{amsmath}
\usepackage{amssymb}
\usepackage{subfigure}
\usepackage{epsfig}
\usepackage{graphicx}
\usepackage{ulem}
\usepackage{color}
\newcommand{\bqa}{\begin{eqnarray}}
\newcommand{\eqa}{\end{eqnarray}}
\newcommand{\beq}{\begin{equation}}
\newcommand{\eeq}{\end{equation}}
\allowdisplaybreaks[1]

\begin{document}
\title{\Large Current-dependent SU(3)-flavor breaking in hidden-heavy $0^{--}$ tetraquark spectroscopy\\[7mm]}

\author{Bing-Dong Wan$^{1,2}$\footnote{wanbd@lnnu.edu.cn}, Ming-Yang Yuan$^{1,2}$, Si-Yu Liu$^{1,2}$, Jun-Hao Zhang$^{1,2}$, and Yan Zhang$^{1,2}$ \vspace{+3pt}}

\affiliation{$^1$Department of Physics, Liaoning Normal University, Dalian 116029, China\\
$^2$Center for Theoretical and Experimental High Energy Physics, Liaoning Normal University, Dalian 116029, China
}

\author{~\\~\\}

\begin{abstract}
\vspace{0.3cm}
We investigate SU(3)-flavor-breaking effects in hidden-heavy tetraquark channels with the exotic quantum number $J^{PC}=0^{--}$, focusing on the dependence of the extracted mass shifts on the interpolating-current structure. Using four local color-octet--octet currents within QCD sum rules, we retain the explicit strange-quark mass and strange condensate contributions in the operator product expansion through dimension eight. By comparing the strange and nonstrange sectors within the same current basis, we obtain hidden-charm-strange mass estimates between 4.37 and 4.79 GeV and hidden-bottom-strange masses between 10.85 and 11.19 GeV. The central charm-sector mass shifts are approximately 40 and 70 MeV for the hidden-flavor-like currents, compared with 190 and 220 MeV for the open-flavor-like currents. The bottom sector exhibits a smaller spread in the corresponding shifts. These results suggest a current-dependent SU(3)-breaking response within the adopted sum-rule framework. The predicted mass associated with one open-flavor-like current also overlaps the $D_s^*\bar D_{s1}(2536)$ threshold region within uncertainties. Our analysis provides complementary strange-sector spectral information while leaving the physical-state composition unresolved.
 \end{abstract}
 \maketitle
\newpage

\section{Introduction}

The channel $J^{PC}=0^{--}$ cannot be realized by an ordinary neutral quark--antiquark meson, so a convincing signal in this channel would immediately point to nonconventional QCD dynamics. For this reason, the $0^{--}$ channel has repeatedly been used as a testing ground for exotic constructions including hybrids, glue-rich states, and tetraquarks~\cite{Qiao:2014vva,Qiao:2015iea,Wan:2023epq,Tang:2025ept,Huang:2016rro}. A natural way to examine such a channel is to modify the flavor content of the light degrees of freedom while keeping the color and Dirac structure of the interpolating field fixed. The replacement $q\to s$ introduces explicit chiral-symmetry breaking, modifies the strange condensates relative to the nonstrange ones, and rearranges nearby hadronic thresholds. In approaches based on operator expansions, such as QCD sum rules~\cite{Shifman:1978bx,Reinders:1984sr}, these changes reorganize the balance among perturbative, mass-suppressed, and condensate-driven contributions. SU(3)-flavor breaking therefore offers a controlled comparison between the strange and nonstrange versions of the same current.

This comparison is especially informative for unconventional hadrons. The modern heavy-hadron spectrum contains many candidates whose quantum numbers or decay properties resist a simple quarkonium interpretation~\cite{Choi:2003ue,Chen:2016qju,Liu:2019zoy,Esposito:2016noz,Guo:2017jvc,Brambilla:2019esw,Chen:2022asf,Wang:2025sic,Zhang:2025qmg}. For such states, the central issue is not only whether they are exotic, but which microscopic degrees of freedom dominate them. In our previous study~\cite{Wan:2026plq}, four local color-octet--octet interpolating currents were employed to investigate the nonstrange hidden-heavy $0^{--}$ channels. The resulting diagonal sum rules provided mass estimates for the hidden-charm and hidden-bottom sectors. An important question left open by that analysis is how the same currents respond when the light quarks are replaced by strange quarks. In particular, it remains to be examined whether the corresponding mass shifts are approximately common to the four currents or depend on their Dirac and flavor arrangements.

Hidden strangeness provides a particularly clean realization of these ideas. Replacing a nonstrange light quark by an $s$ quark introduces explicit $m_s$ terms, replaces nonstrange condensates by strange ones, and rearranges nearby open-flavor thresholds into channels containing $D_s^{(*)}$ mesons. The hidden-charm-strange sector is therefore a natural environment in which to ask whether SU(3)-flavor breaking merely generates an approximately common mass shift or instead differentiates the octet current basis itself. Here ``open-flavor-like'' refers only to the heavy-strange bilinear clustering $[\bar c s]\otimes[\bar s c]$ inside an overall hidden-flavor $c\bar c s\bar s$ system, as opposed to the hidden-flavor-like clustering $[\bar c c]\otimes[\bar s s]$.

Motivated by these considerations, we investigate hidden-charm-strange $0^{--}$ tetraquarks in an octet--octet description. We consider two classes of local operators, $[\bar{c}c]_{8_c}\otimes[\bar{s}s]_{8_c}$ and $[\bar{c}s]_{8_c}\otimes[\bar{s}c]_{8_c}$, and analyze the associated two-point functions within the QCD sum-rule formalism~\cite{Shifman:1978bx,Reinders:1984sr}. The calculation keeps the explicit strange-mass terms and strange condensates through dimension eight, so the hidden-charm-strange system is treated as a dedicated SU(3)-breaking extension of the nonstrange analysis. The hidden-charm-strange sector is the primary phenomenological target, while the hidden-bottom-strange sector is retained as a stability benchmark. In this way, the present work complements earlier studies of hidden-heavy $0^{--}$ systems~\cite{Wan:2026plq,Wan:2024fam} by asking whether the same current basis responds uniformly to the $q\to s$ replacement or whether the induced shifts depend on the selected interpolating current. The comparison is performed at the level of diagonal two-point sum rules. Since the local currents are not unique projectors onto distinct physical configurations, differences in their extracted masses or SU(3)-breaking shifts should not be interpreted as direct measurements of the internal composition of individual states. QCD sum rules remain a standard nonperturbative tool for such questions and have been applied extensively to both conventional and exotic hadrons~\cite{Wan:2026plq,Wan:2024fam,Tang:2024zvf,Tang:2024kmh,Albuquerque:2013ija,P.Col,Narison:1989aq,Govaerts:1984hc,Wang:2013vex,Wan:2019ake,Tang:2021zti,Wan:2020oxt,Wan:2020fsk,Wang:2020cme,Yang:2020wkh,Wang:2021qmn,Wan:2021vny,Yin:2021cbb,Wan:2022xkx,Zhang:2022obn,Wan:2022uie,Agaev:2022pis,Zhao:2023imq,Li:2024ctd,Wan:2024dmi,Wan:2024pet,Wan:2024ykm,Zhang:2024ick,Zhang:2024jvv,Zhang:2024ulk,Zhang:2024asb,Wan:2025xhf,Wan:2025bdr,Wan:2025zau,Agaev:2025llz,Barsbay:2025vjq,Wan:2025fyj,Zhang:2025vqg,Wan:2025sae,Wan:2025ikc,Mu:2026bue,Ben:2025wqn,Tang:2019nwv,Tang:2016pcf,Mutuk:2022ckn,Mu:2026qhw}.

The paper is organized as follows. In Sec.~\ref{Formalism}, we present the QCD sum rule formalism and construct the interpolating currents with color-octet configurations. The numerical analysis and the extraction of the masses are performed in Sec.~\ref{Numerical}. In Sec.~\ref{SU3break}, we analyze the SU(3)-flavor-breaking pattern implied by the strange-sector estimates. Finally, a brief summary is given in Sec.~\ref{Summary}.

\section{Formalism}\label{Formalism}

To isolate the effect of replacing a nonstrange quark by a strange one, we keep the color and Dirac structures fixed and work with a common octet-current basis throughout the analysis. At the quark level, this basis separates naturally into hidden-flavor-like structures, $[\bar{Q}Q]_{\mathbf{8}}\otimes[\bar{s}s]_{\mathbf{8}}$, and open-flavor-like structures, $[\bar{Q}s]_{\mathbf{8}}\otimes[\bar{s}Q]_{\mathbf{8}}$. The four interpolating currents are chosen as
\begin{eqnarray}
J_A(x) &=& i[\bar{Q}_i \gamma_\mu (t^a)_{ij} Q_j][\bar{s}_m \gamma^\mu\gamma_5 (t^a)_{mn} s_n] \;, \\
J_B(x) &=& i[\bar{Q}_i \gamma_\mu\gamma_5 (t^a)_{ij} Q_j][\bar{s}_m \gamma^\mu (t^a)_{mn} s_n] \;, \\
J_C(x) &=& \frac{i}{\sqrt{2}}\Big( [\bar{Q}_i \gamma_\mu (t^a)_{ij} s_j][\bar{s}_m \gamma^\mu\gamma_5 (t^a)_{mn} Q_n] \nonumber \\
&& + [\bar{Q}_i \gamma_\mu\gamma_5 (t^a)_{ij} s_j][\bar{s}_m \gamma^\mu (t^a)_{mn} Q_n] \Big) \;, \\
J_D(x) &=& \frac{i}{\sqrt{2}}\Big( [\bar{Q}_i \gamma_5 (t^a)_{ij} s_j][\bar{s}_m (t^a)_{mn} Q_n] \nonumber \\
&& - [\bar{Q}_i (t^a)_{ij} s_j][\bar{s}_m \gamma_5 (t^a)_{mn} Q_n] \Big) \;.
\end{eqnarray}
In these expressions, the strange field is written explicitly as $s$, and $t^a=\lambda^a/2$ are the $SU(3)_c$ generators. The indices $i$, $j$, $m$, and $n$ label color. The primary analysis concerns the hidden-charm-strange case $Q=c$, while the hidden-bottom-strange benchmark is obtained from the same basis by replacing $Q$ with $b$. Keeping the Dirac and color structures unchanged allows us to compare the strange and nonstrange correlators within the same current basis. The extracted mass shifts, however, reflect both the modified OPE and the associated sum-rule parameter choices. We therefore keep the labels $J_A$, $J_B$, $J_C$, and $J_D$ throughout.

The four currents are representative operators in a selected local color-coupling basis. Their diagonal sum rules do not imply four independent physical states, since local four-quark operators can be related by Fierz transformations~\cite{Wan:2026plq}.

\begin{figure}[htb]
\centering
\includegraphics[width=10cm]{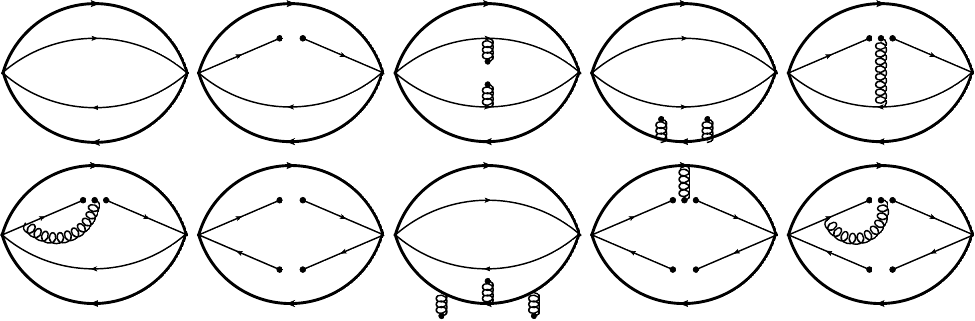}
\caption{Feynman diagrams contributing to the OPE calculation. The thick solid line represents the heavy quark, the thin solid line corresponds to the strange quark, and the coiled line represents the gluon. Contributions from heavy quark condensates are suppressed in the large mass limit.} \label{feyndiag}
\end{figure}

The hadronic information associated with each current is encoded in the standard two-point correlator,
\begin{eqnarray}
\Pi(q^2) &=& i \int d^4 x\, e^{i q \cdot x} \langle 0 | T \{ J(x)\, J^\dagger(0) \} |0 \rangle \;,
\end{eqnarray}
which connects the short-distance quark--gluon dynamics to the hadronic spectral function. In the Euclidean region, we evaluate this object through the operator product expansion and organize the result in spectral form,
\begin{eqnarray}
\Pi^{\mathrm{OPE}}(q^2)
&=& \int_{(2m_Q+2m_s)^2}^{\infty} ds\, \frac{\rho^{\mathrm{OPE}}(s)}{s-q^2}  \;.
\end{eqnarray}
In the strange sector, the spectral density contains not only the perturbative term and the usual condensate contributions through dimension eight, but also explicit $m_s$ corrections distributed among these terms together with the replacements of nonstrange condensates by strange ones. Since the channel-dependent analytical expressions are lengthy, we summarize the OPE content schematically, including the mixed-condensate and higher-dimensional strange-sector terms up to dimension eight, in abbreviated form
\begin{eqnarray}
\rho^{\mathrm{OPE}}(s) &=& \rho^{\mathrm{pert}}(s) +  \rho^{\langle \bar{s}s \rangle}(s) + \rho^{\langle G^2 \rangle}(s) + \dots + \rho^{\langle \bar{s}s \rangle \langle \bar{s}Gs \rangle}(s) \;.
\end{eqnarray}
Representative diagrams contributing to these terms are shown in Fig.~\ref{feyndiag}. We do not reproduce the full expressions here. The key point is that, relative to the nonstrange case, the strange-sector OPE is modified not only by numerical changes in the condensates, but also by explicit $m_s$-weighted terms that rearrange the relative size of different contributions.

On the hadronic side, we separate the lowest hidden-charm-strange pole from the higher resonances and continuum in the usual narrow-resonance-plus-continuum ansatz,
\begin{eqnarray}
\Pi^{\mathrm{phen}}(q^2)
&=& \frac{\lambda_H^2}{M_H^2 - q^2}
+ \int_{s_0}^{\infty} ds\, \frac{\rho^{\mathrm{cont}}(s)}{s - q^2} \;,
\end{eqnarray}
where $M_H$ denotes the tetraquark mass and $\lambda_H$ is the corresponding current--hadron coupling.

After imposing quark--hadron duality and carrying out the Borel transform $\mathcal{B}_{M_B^2}$, the continuum contribution is exponentially suppressed and the working sum rules follow. The hadron mass is extracted from the ratio of Borel moments,
\begin{eqnarray}
M_H(s_0, M_B^2) &=& \sqrt{\frac{\mathcal{L}_1(s_0, M_B^2)}{\mathcal{L}_0(s_0, M_B^2)}} \;,
\end{eqnarray}
where
\begin{eqnarray}
\mathcal{L}_0(s_0, M_B^2) &=& \int_{(2m_Q+2m_s)^2}^{s_0} ds\, \rho^{\mathrm{OPE}}(s) e^{-s/M_B^2}  \;, \\
\mathcal{L}_1(s_0, M_B^2) &=& \frac{\partial}{\partial \left( -1/M_B^2 \right)} \mathcal{L}_0(s_0, M_B^2) \;.
\end{eqnarray}
The role of the Borel parameter $M_B^2$ is twofold: it improves the behavior of the truncated OPE and at the same time weakens the contamination from excited and continuum states. In the present context, it also provides a practical way to see whether the strange-sector corrections destabilize or preserve the windows found in the nonstrange analysis.

\section{Numerical analysis}\label{Numerical}

The numerical input is chosen so as to make the strange-sector effects explicit rather than implicit~\cite{Albuquerque:2013ija,P.Col,Narison:1989aq,Govaerts:1984hc,Tang:2019nwv}. We use $m_c(m_c)=\overline{m}_c=(1.275\pm0.025)\,\text{GeV}$, $m_b(m_b)=\overline{m}_b=(4.18\pm0.03)\,\text{GeV}$, $m_s(2~\mathrm{GeV})=\overline{m}_s=(0.095\pm0.005)\,\text{GeV}$, $\langle \bar{q} q \rangle = -(0.23\pm0.03)^3\,\text{GeV}^3$, $\langle \bar{s} s \rangle = \kappa\,\langle \bar{q} q \rangle$ with $\kappa=0.8\pm0.1$, $\langle \bar{s} g_s \sigma \cdot G s \rangle = m_0^2 \langle\bar{s} s \rangle$, $\langle g_s^2 G^2 \rangle = (0.88\pm0.25)\,\text{GeV}^4$, $\langle g_s^3 G^3 \rangle = (0.045\pm0.013)\,\text{GeV}^6$, and $m_0^2=(0.8\pm0.2)\,\text{GeV}^2$. Unlike the nonstrange analysis, the strange mass is never dropped, and the uncertainties related to $m_s$ and the ratio $\kappa=\langle \bar{s}s \rangle/\langle \bar{q}q \rangle$ are treated as part of the quoted error budget.

For each current, the acceptable Borel window is fixed by the usual compromise between OPE convergence and pole dominance, but the hidden-charm-strange problem gives this procedure a more specific interpretation: one must check whether the SU(3)-breaking terms remain under control while still leaving a sizable pole piece. We monitor the convergence condition through
\begin{eqnarray}
R^{\mathrm{OPE}}=\left|\frac{L_0^{\mathrm{dim}=8}(s_0,M_B^2)}{L_0(s_0,M_B^2)}\right| \; .
\end{eqnarray}
Within the working windows, the dimension-eight term is required to stay subleading compared with the full OPE. In practice, we follow the standard criterion that this contribution remain modest~\cite{P.Col}. The second condition concerns the relative weight of the lowest pole,
\begin{eqnarray}
R^{\mathrm{PC}}=\frac{L_0(s_0,M_B^2)}{L_0(\infty,M_B^2)} \; . \label{RatioPC}
\end{eqnarray}
and we require it to exceed $50\%$. The continuum threshold $s_0$ is then determined together with the Borel interval. For each current, we vary $\sqrt{s_0}$ until the convergence and pole criteria are both satisfied. The lower edge of the admissible window is therefore controlled by the OPE, while the upper edge is set by pole dominance. Among the surviving choices, we adopt the threshold for which the extracted mass displays the mildest Borel dependence. The uncertainty associated with $s_0$ is estimated by varying $\sqrt{s_0}$ by about $0.1~\mathrm{GeV}$ around the central choice~\cite{Wan:2019ake,Wan:2020oxt,Wan:2020fsk}.

The numerical analysis is organized around the quantities that best expose the SU(3)-breaking pattern. These are the mass shift $\Delta M=M_X^{(s)}-M_X^{(q)}$, the change in the preferred continuum threshold, and the displacement of the admissible Borel window relative to the nonstrange baseline~\cite{Wan:2026plq}. The strange-sector results are therefore most informative when read against their nonstrange counterparts. The issue is not merely whether all masses increase after $q\to s$, but whether the increase is approximately common to the four currents or instead depends on the selected current structure. Within this setup, the hidden-charm-strange sector supplies the primary phenomenological signal, whereas the hidden-bottom-strange sector serves as a smoother control sample. In the discussion below, the quoted $\Delta M$ values are used primarily as comparative indicators based on the central mass estimates, and the finer hierarchy among them should be interpreted together with the quoted mass uncertainties.

\begin{figure}
\includegraphics[width=6.8cm]{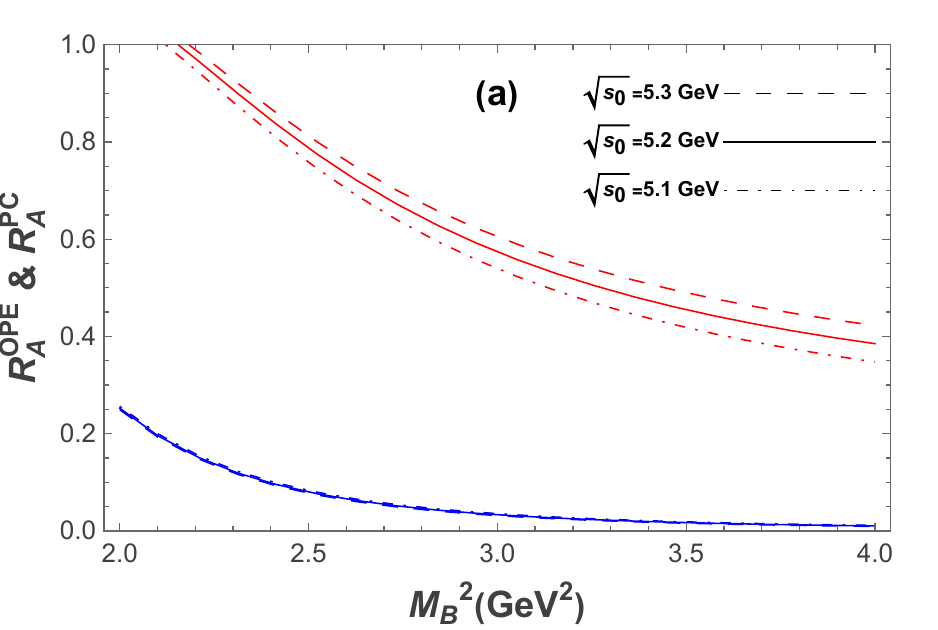}
\includegraphics[width=6.8cm]{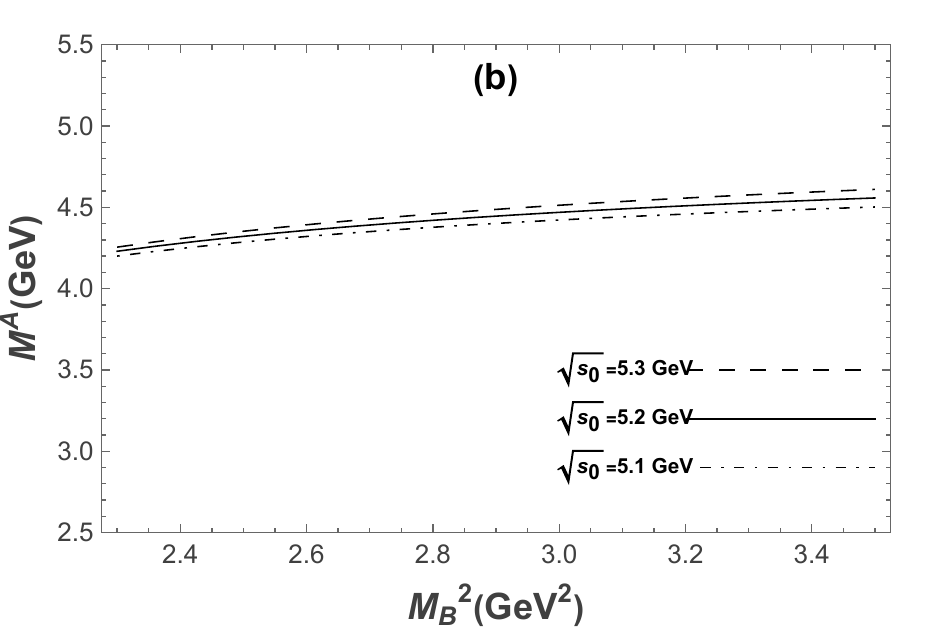}
\caption{(a) The ratios of ${R_{A}^{OPE}}$ and ${R_{A}^{PC}}$ as functions of the Borel parameter $M_B^2$ for different values of $\sqrt{s_0}$, where blue lines represent ${R_{A}^{OPE}}$ and red lines denote ${R_{A}^{PC}}$. (b) The mass $M_{A}^{cs}$ as a function of the Borel parameter $M_B^2$ for different values of $\sqrt{s_0}$.} \label{figAmain}
\end{figure}

We begin with the hidden-charm-strange sector, which is the primary phenomenological target of the present work. For the current $J_A$, the Borel behavior of the OPE ratio, the pole contribution, and the extracted mass are displayed in Fig.~\ref{figAmain}. The figure shows that the highest-dimension term remains numerically suppressed inside the selected interval, while the pole contribution decreases monotonically with $M_B^2$ and is the quantity that sets the upper edge of the admissible window. Because the strange pieces alter the Wilson coefficients, the preferred window and threshold need not coincide with those of the nonstrange study. For the representative central choice $\sqrt{s_0}=(5.2\pm0.1)~\mathrm{GeV}$, our strange-sector result gives
\begin{eqnarray}
2.4 \le M_B^2 \le 3.3~\text{GeV}^2 \; ,
\end{eqnarray}
which leads to the hidden-charm-strange mass estimate
\begin{eqnarray}
M_{A}^{cs}=(4.42\pm0.15)\,\text{GeV} \; .
\end{eqnarray}
This channel therefore remains compatible with a low-lying hidden-charm-strange exotic configuration of quantum numbers $J^{PC}=0^{--}$ even after SU(3)-breaking effects are taken into account. Once the corresponding nonstrange benchmark~\cite{Wan:2026plq} is placed alongside it, the quantity of immediate interest is not only the central mass itself but also the induced shift, here about $40~\mathrm{MeV}$, and whether the strange corrections preserve the Borel stability of the hidden-flavor-like current. At the level of the central estimates, the $J_A$ channel exhibits a relatively small strange--nonstrange mass shift. This property of the corresponding diagonal sum rules does not by itself establish a rigid or compact physical configuration.

For the remaining hidden-charm-strange currents, the same broad pattern persists, although with stronger current dependence. For $J_B$, with the representative central choice $\sqrt{s_0}=(5.0\pm0.1)~\mathrm{GeV}$, we obtain $(4.37\pm0.10)$ GeV in the window $2.1\le M_B^2\le2.7~\text{GeV}^2$. For $J_C$, the preferred strange-sector threshold is raised to $\sqrt{s_0}=(5.6\pm0.1)~\mathrm{GeV}$ and the admissible Borel interval broadens to $3.0\le M_B^2\le4.0~\text{GeV}^2$, leading to the mass estimate $(4.79\pm0.13)$ GeV. For $J_D$, with $\sqrt{s_0}=(5.4\pm0.1)~\mathrm{GeV}$ and $2.6\le M_B^2\le3.3~\text{GeV}^2$, we obtain $(4.65\pm0.11)$ GeV. In the charm-strange sector, the central masses of the hidden-flavor-like currents remain close to one another, whereas those of the open-flavor-like currents lie higher. This is the ordering suggested by the present central estimates, with $J_A$ and $J_B$ remaining close, $J_D$ occupying an intermediate position, and $J_C$ staying highest among the four currents. The $J_C$ channel requires the highest strange-sector continuum threshold and the widest admissible Borel interval among the four charm currents, while the $J_D$ central mass lies in the vicinity of the $D_s^*\bar D_{s1}$ threshold region.

The hidden-bottom-strange sector is retained as a control benchmark. For the same current $J_A$, the Borel analysis remains comparatively stable. For the representative threshold choice $\sqrt{s_0}=(11.6\pm0.1)~\mathrm{GeV}$, we obtain
\begin{eqnarray}
M_{A}^{bs}=(10.85\pm0.09)\,\text{GeV} \; ,
\end{eqnarray}
in the working interval
\begin{eqnarray}
10.4 \le M_B^2 \le 11.5~\text{GeV}^2 \; .
\end{eqnarray}
The bottom-strange channel displays a flatter Borel platform and a milder strange-sector shift, which makes it a useful reference benchmark rather than the main phenomenological target. For the remaining bottom-strange currents, we obtain $(10.89\pm0.09)$ GeV for $J_B$, $(11.19\pm0.09)$ GeV for $J_C$, and $(11.01\pm0.09)$ GeV for $J_D$, with Borel windows $10.1\le M_B^2\le11.0~\text{GeV}^2$, $10.7\le M_B^2\le11.9~\text{GeV}^2$, and $10.4\le M_B^2\le11.5~\text{GeV}^2$, respectively. The corresponding representative thresholds are $\sqrt{s_0}=(11.6\pm0.1)~\mathrm{GeV}$ for $J_B$, $\sqrt{s_0}=(12.1\pm0.1)~\mathrm{GeV}$ for $J_C$, and $\sqrt{s_0}=(11.8\pm0.1)~\mathrm{GeV}$ for $J_D$. Compared with the charm sector, the bottom-strange estimates vary more smoothly and their SU(3)-breaking shifts show a smaller spread.

The combined pattern of the charm and bottom results should be read as a paired comparison rather than as isolated spectra. The central question is whether the substitution $q\to s$ produces approximately uniform shifts or instead differentiates the four current structures. In the central estimates, the charm sector shows a broader separation, where the hidden-flavor-like currents $J_A$ and $J_B$ remain less shifted than the open-flavor-like currents $J_C$ and $J_D$, while the bottom sector supplies a smoother control sample.

Viewed globally, the mass spectrum is ordered rather than random. The hidden-flavor-like octet currents $J_A$ and $J_B$, corresponding to the structures $[\bar c c]_{8_c}\otimes[\bar s s]_{8_c}$, generally lie below the open-flavor-like currents $J_C$ and $J_D$, built from the structures $[\bar c s]_{8_c}\otimes[\bar s c]_{8_c}$, with $J_D$ sitting between the hidden-flavor-like pair and $J_C$. At the same time, the four current-associated estimates remain concentrated within a relatively narrow mass band. For the strange system, the more interesting question is whether SU(3) breaking preserves this ordering or distorts it. The most economical way to answer that question is to compare, current by current, the signs and magnitudes of $\Delta M$, the shifts in $\sqrt{s_0}$, and the changes in the Borel windows. In the present charm-sector central estimates, the strange corrections preserve the ordering $M_A^{cs}\simeq M_B^{cs}<M_D^{cs}<M_C^{cs}$ and suggest a pattern in which the hidden-flavor-like currents remain less shifted than the open-flavor-like ones. The uncertainties of the individual mass estimates are, however, substantial, and the significance of this separation cannot be established from the central values alone. A more complete discussion of current mixing deserves a dedicated follow-up analysis.

Another point worth emphasizing is the possible proximity of the charm-strange central masses from $J_C$ and $J_D$ to nearby open-strange thresholds such as $D_{s1}\bar D_s^*$~\cite{Peng:2022nrj}. For orientation, the representative thresholds are about $2.112+2.460\simeq4.57~\mathrm{GeV}$ for $D_s^*\bar D_{s1}(2460)$ and $2.112+2.536\simeq4.65~\mathrm{GeV}$ for $D_s^*\bar D_{s1}(2536)$~\cite{ParticleDataGroup:2024cfk}. The predicted charm-strange masses can therefore be compared with these open-strange thresholds. In particular, the central mass extracted from $J_D$, $4.65\pm0.11$ GeV, overlaps the $D_s^*\bar D_{s1}(2536)$ threshold within uncertainties, while the $J_C$ central value lies higher. This proximity identifies a potentially relevant energy region for experimental searches. It does not, however, establish an enhanced meson--meson component, a threshold-induced mass shift, or a more compact internal configuration. Such conclusions would require dynamical information beyond the diagonal two-point sum rules considered here.

Overall, both the bottom-strange and charm-strange estimates lie in narrow mass regions. The physically relevant contrast between them is less the raw mass spread than the stability of the Borel platforms and the visibility of the SU(3)-breaking corrections. The bottom-strange channels display flatter mass curves and hence a more stable extraction, whereas the charm-strange channels react more strongly to changes in the Borel window and threshold. The representative windows adopted for the four currents are those listed in Table~\ref{mass}, with the lower edge controlled by OPE convergence and the upper edge by pole dominance. The numerical results are summarized in Table~\ref{mass}, while the Borel stability, OPE-convergence, and pole-contribution curves are shown in Fig.~\ref{figAmain} and Appendix~\ref{pictures}. In this setup, the hidden-bottom-strange channels provide a control sample for judging which features are generic heavy-quark effects and which are specifically amplified in the charm sector.

For convenience, the outputs for all four currents are collected in Table~\ref{mass}. To make the SU(3)-breaking pattern transparent, the table is arranged so that the nonstrange benchmark, the strange result, and the induced shifts can be compared current by current. In keeping with the main emphasis of the present work, the hidden-charm and hidden-charm-strange comparison is listed first and should be regarded as the primary phenomenological output, while the hidden-bottom and hidden-bottom-strange entries are placed afterward as a control benchmark. The nonstrange $q$-quark baselines collected in Table~\ref{mass} are taken from the color-octet hidden-heavy $0^{--}$ analysis of Ref.~\cite{Wan:2026plq}, while the strange rows are obtained in the present work. The main text uses the hidden-charm-strange $J_A$ channel as the representative example, while the remaining stability plots are deferred to Appendix~\ref{pictures}, where the charm-sector figures are shown before the bottom-sector ones. The quoted errors primarily reflect the variation of the heavy masses, condensates, and continuum thresholds; in the strange-sector analysis, uncertainties tied to the strange inputs also contribute through the SU(3)-breaking pieces of the OPE.

\begin{table}[!htbp]
\begin{center}
\renewcommand\tabcolsep{4pt}
\caption{Minimal SU(3)-flavor-breaking comparison for the $0^{--}$ color-octet tetraquark candidates. For each current, the nonstrange ($q$) and strange ($s$) results are arranged as a two-line block so that the changes in $\sqrt{s_0}$, Borel window, and extracted mass can be read vertically. The induced mass shift $\Delta M$ is listed at the end of the strange row. The hidden-charm sector provides the primary phenomenological comparison, while the hidden-bottom sector is included as a stability benchmark. The quoted mass shifts are differences between central values; correlated uncertainties of the shifts have not been separately determined.}\label{mass}
\resizebox{\textwidth}{!}{%
\begin{tabular}{ccccccc}\hline\hline
Sector & Current & Flavor & $\sqrt{s_0}$ (GeV) & $M_B^2$ (GeV$^2$) & $M_X$ (GeV) & $\Delta M$ (GeV) \\ \hline
$c/c_s$ &$A$ &$q$ & $(5.1\pm0.1)$   & $2.5$-$3.1$   & $(4.38\pm0.13)$ &  \\
        &     &$s$ & $(5.2\pm0.1)$   & $2.4$-$3.3$   & $(4.42\pm0.15)$ & $+0.04$ \\
        &$B$ &$q$ & $(4.9\pm0.1)$   & $2.0$-$2.7$   & $(4.30\pm0.15)$ &  \\
        &     &$s$ & $(5.0\pm0.1)$   & $2.1$-$2.7$   & $(4.37\pm0.10)$ & $+0.07$ \\
        &$C$ &$q$ & $(5.3\pm0.1)$   & $2.6$-$3.3$   & $(4.57\pm0.11)$ &  \\
        &     &$s$ & $(5.6\pm0.1)$   & $3.0$-$4.0$   & $(4.79\pm0.13)$ & $+0.22$ \\
        &$D$ &$q$ & $(5.1\pm0.1)$   & $2.3$-$2.9$   & $(4.46\pm0.11)$ &  \\
        &     &$s$ & $(5.4\pm0.1)$   & $2.6$-$3.3$   & $(4.65\pm0.11)$ & $+0.19$ \\\hline
$b/b_s$ &$A$ &$q$ & $(11.5\pm0.1)$ & $10.2$-$11.2$ & $(10.78\pm0.09)$ &  \\
        &     &$s$ & $(11.6\pm0.1)$ & $10.4$-$11.5$ & $(10.85\pm0.09)$ & $+0.07$ \\
        &$B$ &$q$ & $(11.5\pm0.1)$ & $10.0$-$10.9$ & $(10.82\pm0.08)$ &  \\
        &     &$s$ & $(11.6\pm0.1)$ & $10.1$-$11.0$ & $(10.89\pm0.09)$ & $+0.07$ \\
        &$C$ &$q$ & $(11.9\pm0.1)$ & $10.6$-$11.6$ & $(11.08\pm0.08)$ &  \\
        &     &$s$ & $(12.1\pm0.1)$ & $10.7$-$11.9$ & $(11.19\pm0.09)$ & $+0.11$ \\
        &$D$ &$q$ & $(11.6\pm0.1)$ & $10.2$-$11.2$ & $(10.90\pm0.08)$ &  \\
        &     &$s$ & $(11.8\pm0.1)$ & $10.4$-$11.5$ & $(11.01\pm0.09)$ & $+0.11$ \\
\hline
\hline
\end{tabular}
}
\end{center}
\end{table}
 
\section{SU(3)-Flavor-Breaking Pattern}
\label{SU3break}

The central issue in the present work is not whether the strange replacement shifts all four octet currents upward, but whether it does so in a universal way. A common SU(3)-breaking response would produce nearly equal $\Delta M$ values for $J_A$--$J_D$ and would therefore leave the current basis essentially unchanged apart from an overall offset. The comparison in Table~\ref{mass} indicates that the central SU(3)-breaking mass shifts are not identical among the four interpolating currents. In the charm sector, the shifts associated with $J_A$ and $J_B$ are approximately 40 and 70~MeV, while those associated with $J_C$ and $J_D$ are approximately 220 and 190~MeV, respectively. A smaller separation is observed in the bottom sector. This pattern suggests that the flavor-breaking response of the diagonal sum rules depends on the selected current structure. However, the uncertainties of the individual mass estimates are substantial, and the significance of the separation between the two current groups cannot be established from the central values alone.

The different Dirac and flavor arrangements lead to different Wilson coefficients and spectral weights in the corresponding correlation functions. Consequently, the strange-mass and condensate corrections need not produce identical changes in their Borel moments. The larger central mass shifts for $J_C$ and $J_D$ are also accompanied by larger changes in the preferred continuum thresholds: in the charm sector the change in $\sqrt{s_0}$ is about $+0.3$~GeV for $J_C$ and $J_D$, compared with about $+0.1$~GeV for $J_A$ and $J_B$. The observed pattern therefore reflects the combined OPE and sum-rule parameter dependence. The present calculation does not isolate the individual contributions responsible for the shifts, and the change in the continuum threshold should not be regarded as an independently observable dynamical mechanism.

The predicted charm-strange masses can also be compared with nearby open-strange thresholds. In particular, the mass extracted from $J_D$, $4.65\pm0.11$~GeV, overlaps the $D_s^*\bar D_{s1}(2536)$ threshold within uncertainties. This proximity identifies a potentially relevant energy region for experimental searches. It does not, however, establish an enhanced meson--meson component or a threshold-induced mass shift. Such conclusions would require dynamical information beyond the diagonal two-point sum rules considered here.

In the bottom sector, the rounded central mass shifts cluster around 0.07 and 0.11 GeV, whereas the charm sector exhibits a larger spread. These patterns characterize the central estimates of the selected diagonal sum rules and should not be interpreted as exact degeneracies or statistically resolved hierarchies.

These differences characterize the diagonal sum rules associated with the selected interpolating operators. They do not by themselves establish the number of distinct physical states, their spatial configurations, or the strength of their coupling to specific two-meson channels.

\section{Summary}
\label{Summary}

This work examines hidden-charm-strange tetraquark candidates with $J^{PC}=0^{--}$ in a color-octet framework using QCD sum rules. Because the $0^{--}$ quantum numbers cannot be produced by a conventional neutral quark--antiquark meson, the channel gives access to genuinely exotic dynamics.

Using four local color-octet--octet currents and an OPE truncated at dimension eight, we construct a hidden-charm-strange sum-rule system in which the strange mass and strange condensates remain explicit from start to finish. The resulting mass estimates are summarized in Table~\ref{mass}. Numerically, the hidden-charm-strange channels are more sensitive to the choice of threshold and Borel window than the bottom benchmark; the representative windows for all four currents are those listed in Table~\ref{mass}. At the level of the central estimates, the SU(3)-breaking mass shifts are not uniform across the four interpolating currents. The charm-sector central shifts are approximately 40 and 70~MeV for the hidden-flavor-like currents and approximately 190 and 220~MeV for the open-flavor-like ones, while the bottom sector exhibits a smaller spread in the corresponding shifts. The individual mass uncertainties, together with the choice of continuum threshold, are not sufficient to establish the significance of this separation quantitatively, and the shifts should therefore be read as comparative indicators based on central values within the adopted windows.

The phenomenological implication is that the predicted charm-strange masses lie in a region that can be confronted with open-strange thresholds. In particular, the mass extracted from one open-flavor-like current, $4.65\pm0.11$~GeV, overlaps the $D_s^*\bar D_{s1}(2536)$ threshold within uncertainties, which identifies a potentially relevant energy region for experimental searches. This proximity does not establish an enhanced meson--meson component, a threshold-induced mass shift, or a more compact internal configuration.

The decay phenomenology can be summarized briefly. Because a neutral $0^{--}$ hidden-charm-strange state cannot fall apart into the lightest pseudoscalar pair $D_s\bar D_s$, the more relevant search channels are expected to be $D_s\bar D_s^*$ and $D_s^*\bar D_{s1}$, together with hidden-flavor strange modes such as $J/\psi\,\eta$, $J/\psi\,\eta'$, and $\eta_c\,\phi$. These channels define the mass region in which the predicted states could be sought experimentally.

These results extend the previous nonstrange analysis~\cite{Wan:2026plq} to hidden-strange $0^{--}$ channels and suggest a nonuniform flavor-breaking response among the selected interpolating currents. The corresponding mass estimates provide additional spectral information for future studies, while the identification of distinct physical states and their internal compositions remains beyond the scope of the present diagonal sum-rule analysis.

\vspace{0.5cm} {\bf Acknowledgments}

This work was supported in part by the National Natural Science Foundation of China under Grants 12575106 and 12147214, and Specific Fund of Fundamental Scientific Research Operating Expenses for Undergraduate Universities in Liaoning Province under Grants No. LJ212410165019.
During the preparation of this manuscript, the authors used ChatGPT (OpenAI) only for language polishing. The authors carefully checked and revised the manuscript and take full responsibility for all scientific content, calculations, analyses, interpretations, and conclusions.

\clearpage
\begin{widetext}
\appendix
\section{Supplementary Figures}\label{pictures}
Since the charm-sector current $J_A$ is already shown in the main text as the representative example, the supplementary charm-sector plots are limited to the remaining currents $J_B$-$J_D$. Their OPE, pole contribution, and masses as functions of the Borel parameter $M_B^2$ are given in Figs.~\ref{figB0--} to \ref{figD0--}. For completeness, the corresponding hidden-bottom-strange benchmark curves for $J_A$-$J_D$ are also displayed in Figs.~\ref{figAb0--} to \ref{figDb0--}. 

\begin{figure}[h]
\includegraphics[width=6.8cm]{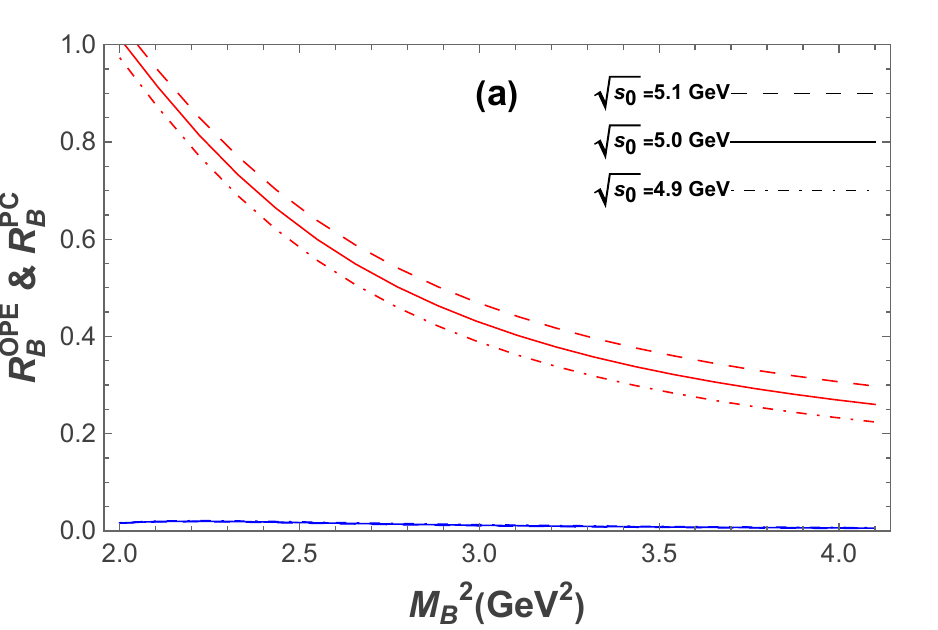}
\includegraphics[width=6.8cm]{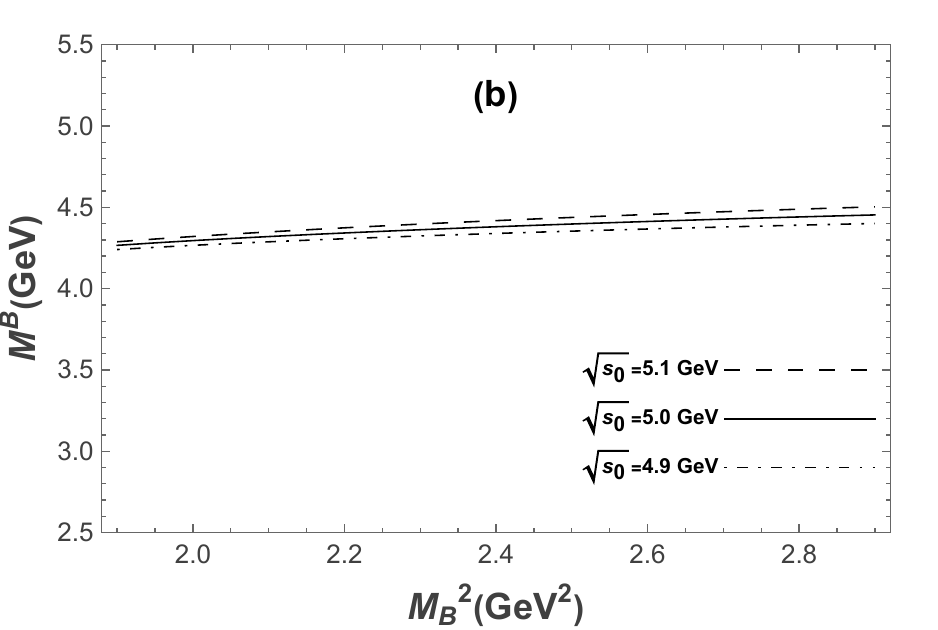}
\caption{OPE ratio, pole contribution, and mass curves for the hidden-charm-strange tetraquark state associated with the current $J_B$.} \label{figB0--}
\end{figure}

\begin{figure}[h]
\includegraphics[width=6.8cm]{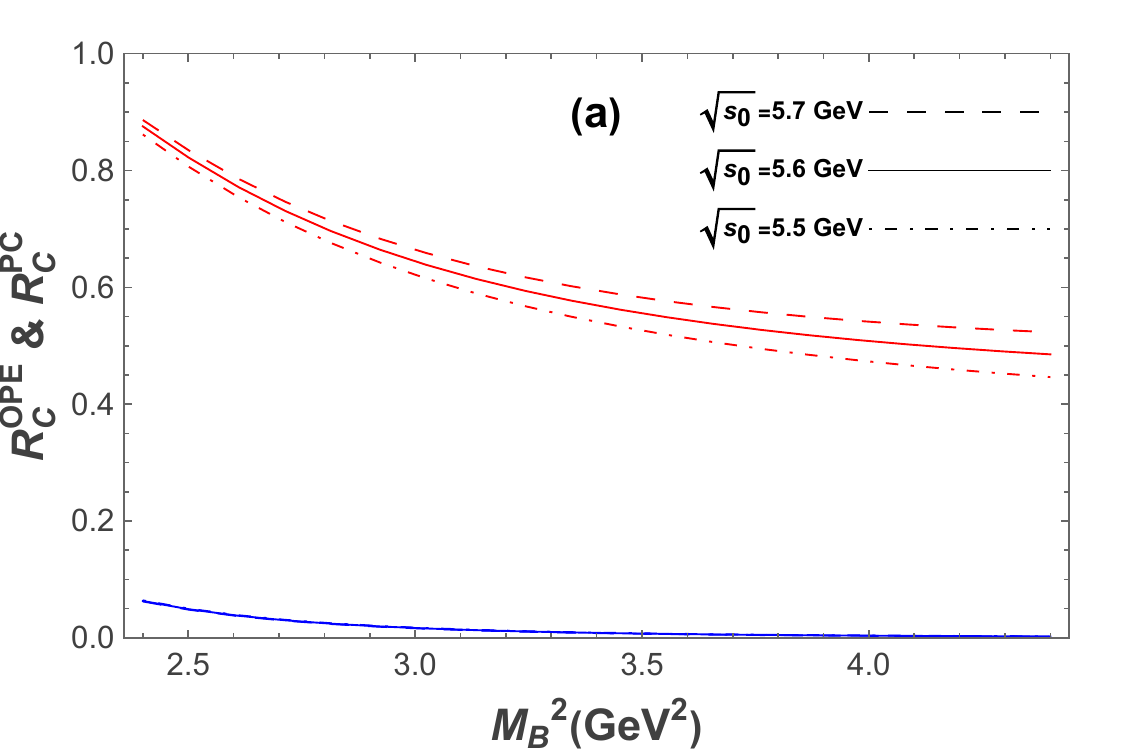}
\includegraphics[width=6.8cm]{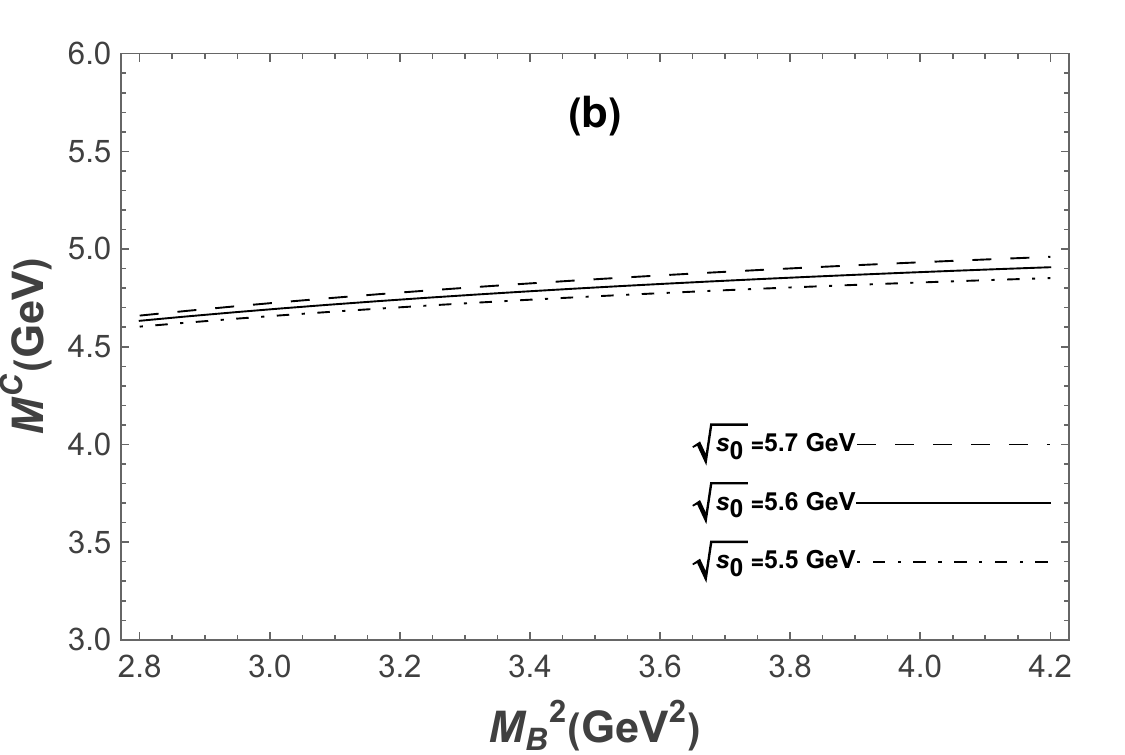}
\caption{OPE ratio, pole contribution, and mass curves for the hidden-charm-strange tetraquark state associated with the current $J_C$.} \label{figC0--}
\end{figure}

\begin{figure}[h]
\includegraphics[width=6.8cm]{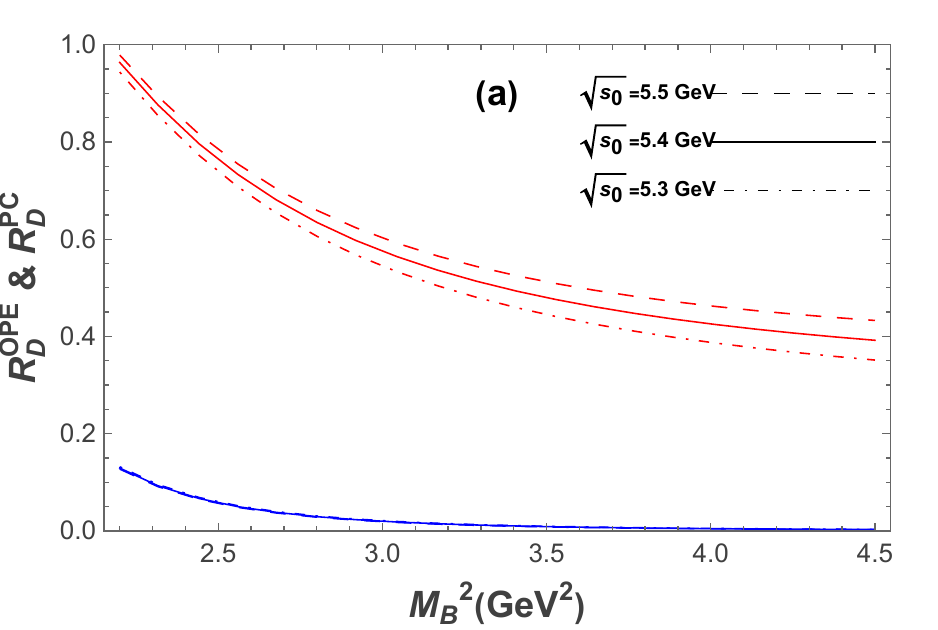}
\includegraphics[width=6.8cm]{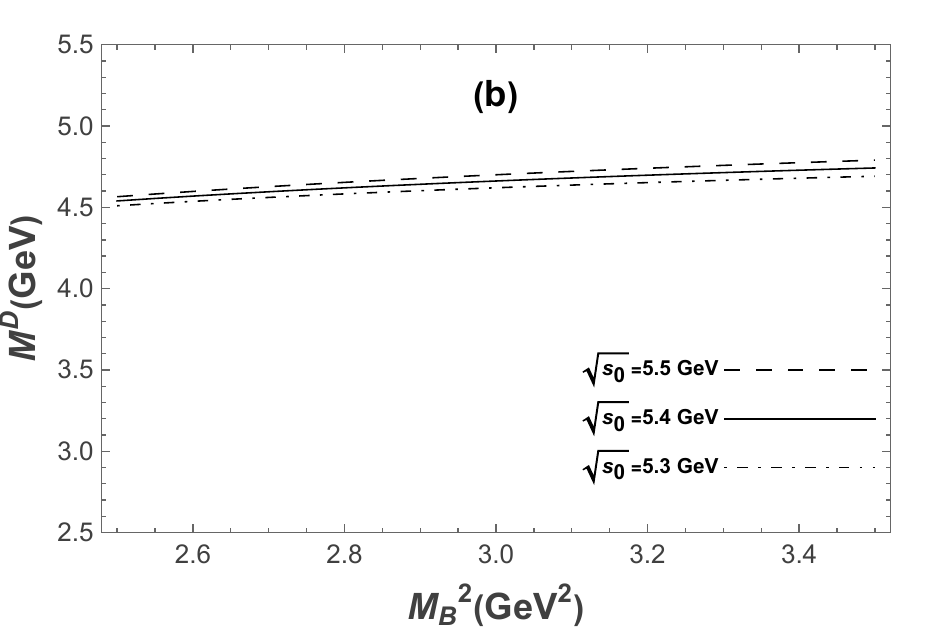}
\caption{OPE ratio, pole contribution, and mass curves for the hidden-charm-strange tetraquark state associated with the current $J_D$.} \label{figD0--}
\end{figure}

\begin{figure}[h]
\includegraphics[width=6.8cm]{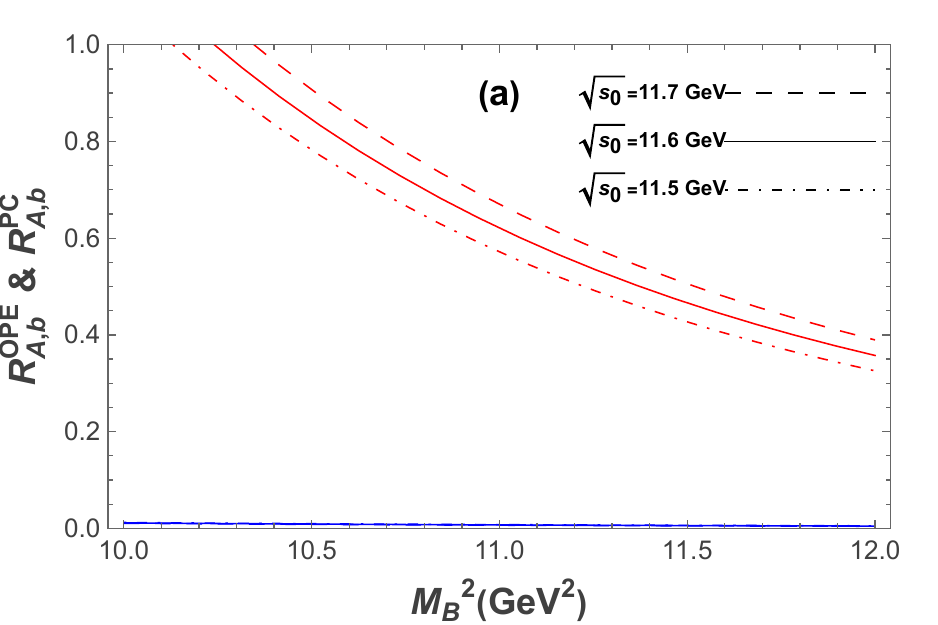}
\includegraphics[width=6.8cm]{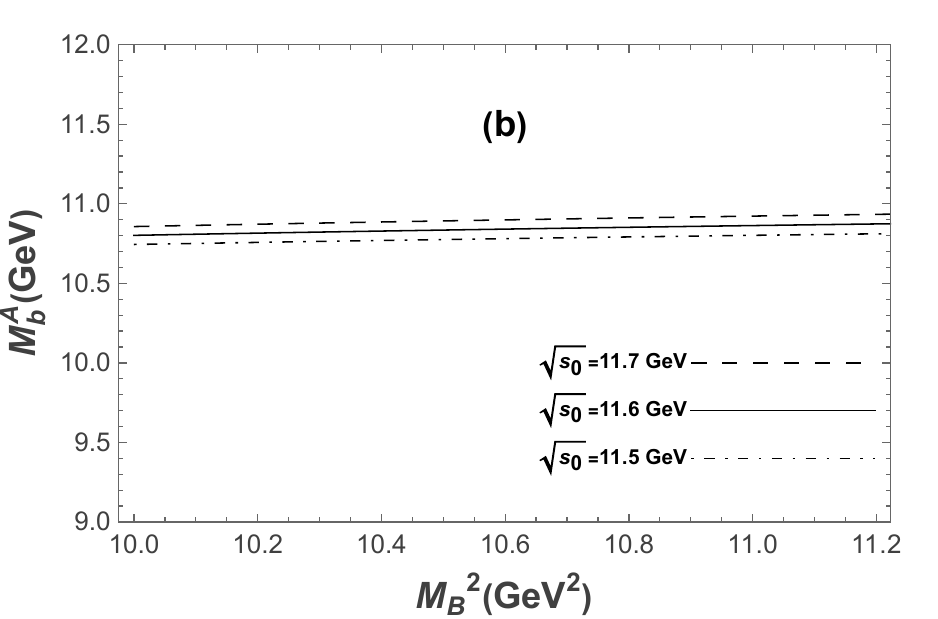}
\caption{OPE ratio, pole contribution, and mass curves for the hidden-bottom-strange benchmark state associated with the current $J_A$.} \label{figAb0--}
\end{figure}

\begin{figure}[h]
\includegraphics[width=6.8cm]{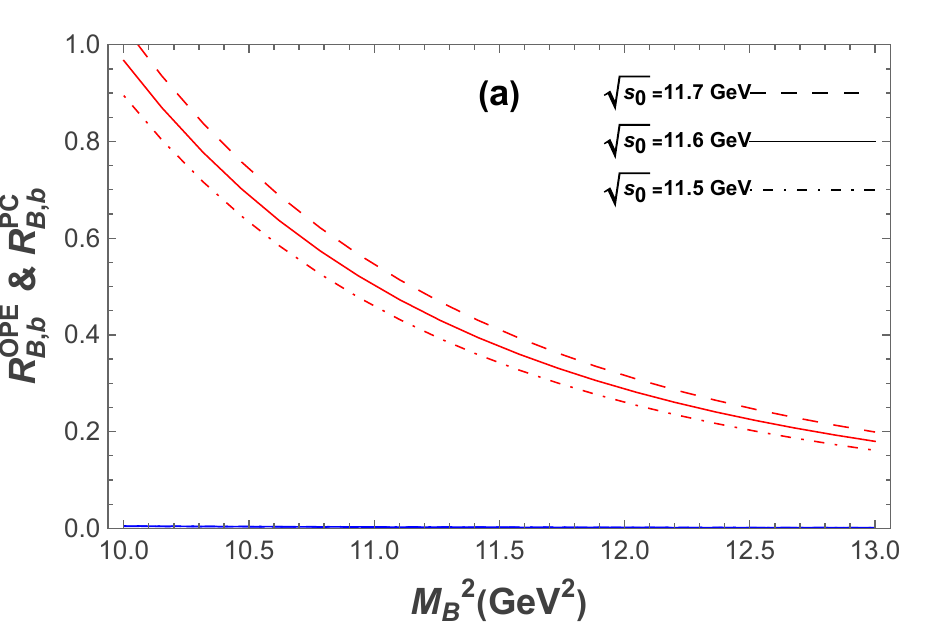}
\includegraphics[width=6.8cm]{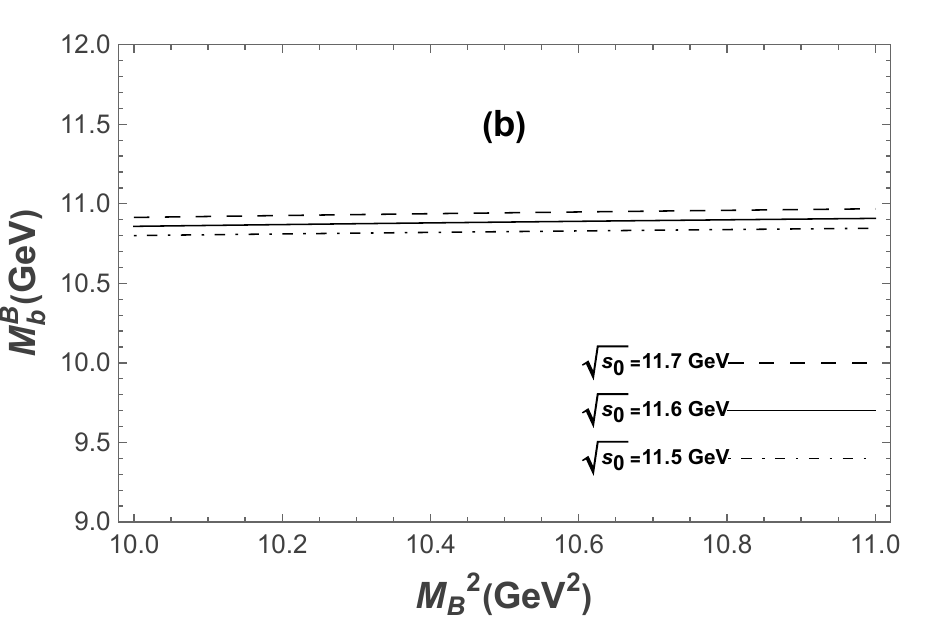}
\caption{OPE ratio, pole contribution, and mass curves for the hidden-bottom-strange benchmark state associated with the current $J_B$.} \label{figBb0--}
\end{figure}

\begin{figure}[h]
\includegraphics[width=6.8cm]{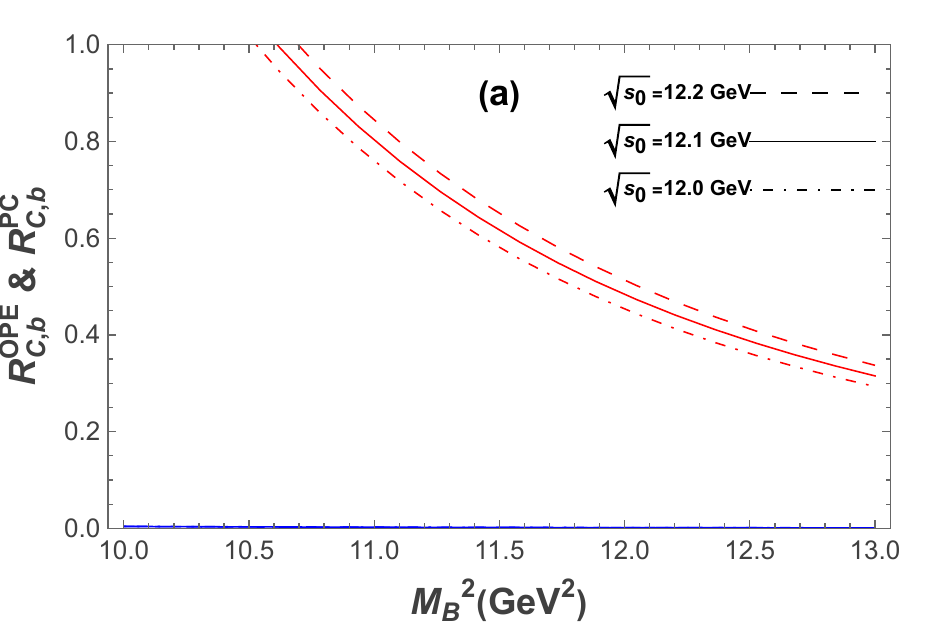}
\includegraphics[width=6.8cm]{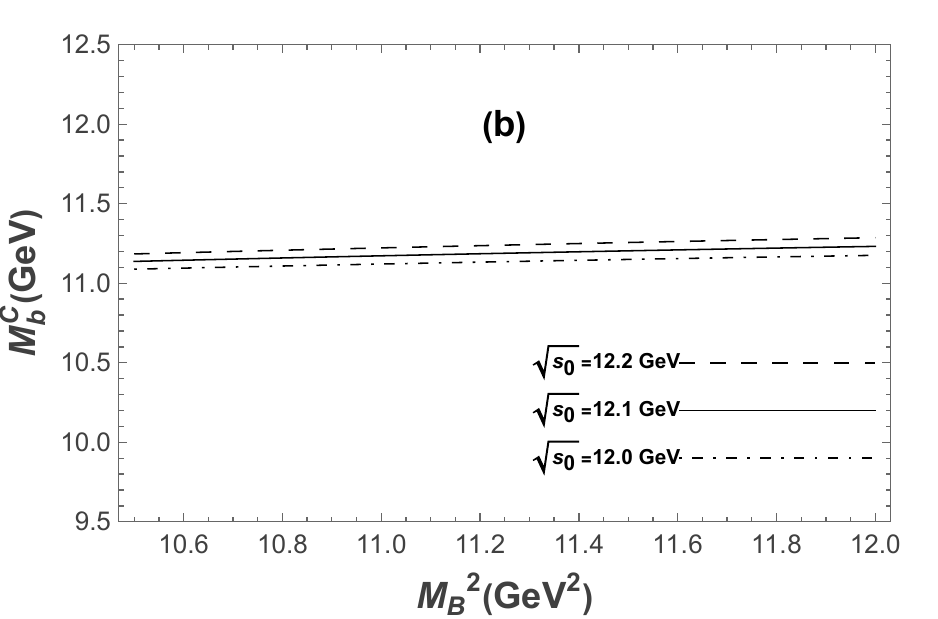}
\caption{OPE ratio, pole contribution, and mass curves for the hidden-bottom-strange benchmark state associated with the current $J_C$.} \label{figCb0--}
\end{figure}

\begin{figure}[h]
\includegraphics[width=6.8cm]{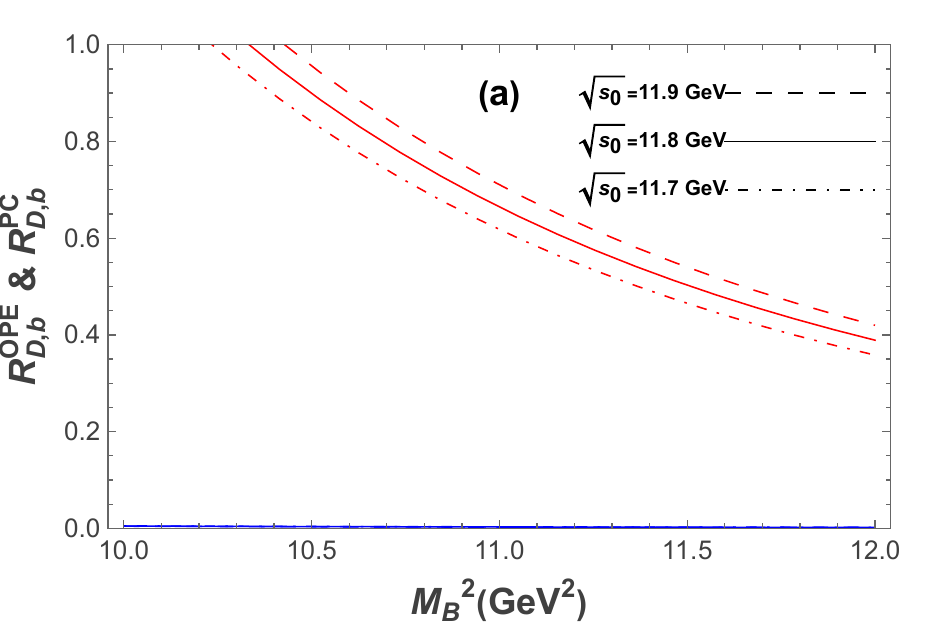}
\includegraphics[width=6.8cm]{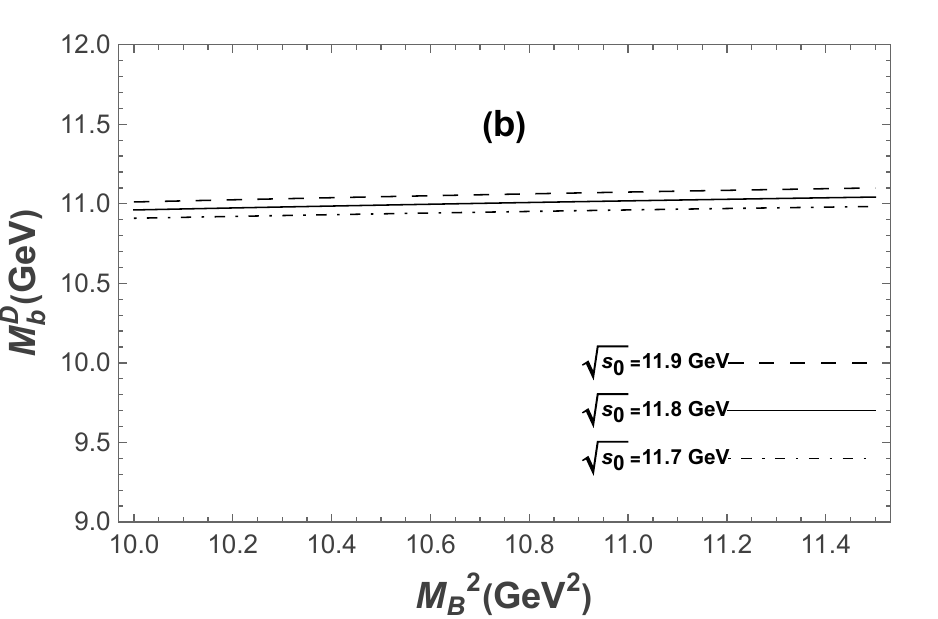}
\caption{OPE ratio, pole contribution, and mass curves for the hidden-bottom-strange benchmark state associated with the current $J_D$.} \label{figDb0--}
\end{figure}

\end{widetext}
\end{document}